\documentclass[a4paper,fleqn,usenatbib]{mnras}
\usepackage{txfonts}
\usepackage[T1]{fontenc}
\usepackage{ae,aecompl}
\usepackage{graphicx}	
\title[MaNGA mass density - metalicity relation]{Do galaxy global relationships emerge from local ones? \\ The SDSS IV MaNGA surface mass density -- metallicity relation}
\author[Jorge K. Barrera-Ballesteros et al.]{
Jorge~K.~Barrera-Ballesteros$^{1}$\thanks{E-mail: jbarrer3@jhu.edu},\, 
Timothy M. Heckman$^{1}$, 
Guangtun~B.~Zhu$^{1}$\thanks{Hubble Fellow}, \and 
Nadia~L.~Zakamska$^{1,2}$, 
Sebastian~F.~S\'{a}nchez$^{3}$, 
David~Law$^{4}$, 
David.~Wake$^{5,6}$, \and
Jenny~E.~Green$^{7}$,
Dmitry~Bizyaev$^{8,9}$,
Daniel~Oravetz$^{8}$,
Audrey~Simmons$^{8}$, \and
Elena~Malanushenko$^{8}$,  
Kaike~Pan$^{8}$,
Alexandre~Roman~Lopes$^{10}$,
Richard~R.~Lane$^{11}$
\\
$^{1}$Department of Physics \& Astronomy, Johns Hopkins University, Bloomberg Center, 3400 N. Charles St., Baltimore, MD 21218, USA \\
$^{2}${Deborah Lunder and Alan Ezekowitz Founders' Circle Member, Institute for Advanced Study, Einstein Dr., Princeton, NJ 08540, USA} \\
$^{3}$Instituto de Astronom\'ia, Universidad Nacional Aut\'onoma de M\'exico, A.P. 70-264, 04510 M\'exico, D.F., M\'exico \\  
$^{4}$Space Telescope Science Institute, 3700 San Martin Drive, Baltimore, MD 21218, USA \\
$^{5}$Department of Physical Sciences, The Open University, Milton Keynes, MK7 6AA, UK. \\
$^{6}$Department of Astronomy, University of Wisconsin-Madison, Madison, WI 53706, USA. \\
$^{7}$Princeton University, Department of Astrophysical Sciences Princeton, NJ 08544, USA \\
$^{8}$Apache Point Observatory and New Mexico State University, P.O. Box 59, Sunspot, NM, 88349-0059, USA \\
$^{9}$Sternberg Astronomical Institute, Moscow State University, Moscow, 119992, Russia. \\
$^{10}$Departamento de F\'{i}sica, Facultad de Ciencias, Universidad de La Serena, Cisternas 1200, La Serena, Chile \\
$^{11}$Instituto de Astrof\'{i}sica,  Pontificia Universidad Cat\'olica de Chile, Av. Vicuna Mackenna 4860, 782-0436 Macul, Santiago, Chile
}

\date{Accepted XXX. Received YYY; in original form ZZZ}
\pubyear{2016}

\begin{document}
\label{firstpage}
\pagerange{\pageref{firstpage}--\pageref{lastpage}}
\maketitle

\begin{abstract}
We present the stellar surface mass density {\it vs.} gas metallicity ($\Sigma_*-Z$) relation for more than 500,000 spatially-resolved star-forming resolution elements (spaxels) from a sample of 653 disk galaxies included in the SDSS IV MaNGA survey. We find a tight relation between these local properties, with higher metallicities as the surface density increases. This relation extends over three orders of magnitude in the surface mass density and a factor of four in metallicity. We show that this local relationship can simultaneously reproduce two well-known properties of disk galaxies: their global mass-metallicity relationship {\it and} their radial metallicity gradients. We also find that the $\Sigma_* - Z$ relation is largely independent of the galaxy's total stellar mass and specific star-formation rate (sSFR), except at low stellar mass and high sSFR. These results suggest that in the present-day universe local properties play a key role in determining the gas-phase metallicity in typical disk galaxies.
\end{abstract}

\begin{keywords}
galaxies: abundances -- galaxies: ISM -- galaxies: fundamental parameters  -- techniques: imaging spectroscopy
\end{keywords}


\section{Introduction}

The observed oxygen abundance in the interstellar medium and the stellar mass of a galaxy are the result of galaxy evolution. The metallicity measured from the ionized gas emission lines reflects the metal content produced in previous generations of stars, while the stellar mass traces the previous gas available for star formation. Therefore, the analysis of how these properties correlate is fundamental to our understanding of the assembly of galaxies across cosmic time. The physical mechanisms explaining the connection between these two fundamental galactic features are a matter of debate. An important clue would be to establish whether these relations are related primarily to the global properties of the galaxy {\it vs.} the local ones. 

There are two key systematic properties that have been uncovered. The first is the close relation between total stellar mass and metallicity. This has been explored for almost 40 years. In a seminal study, \cite{1979A&A....80..155L} found a positive correlation between the oxygen abundance and the luminosity (a proxy for the stellar mass) in a sample of irregular and compact galaxies. Later studies found similar trends for samples of different Hubble types, including a wide range of luminosities and metallicities \cite[e.g,][]{1989ApJ...347..875S, 1994ApJ...420...87Z}. With the development of large spectroscopic surveys such as the single-fiber Sloan Digital Sky Survey (SDSS), \cite{2004ApJ...613..898T} confirmed the mass-metallicity relation (MZR) for a sample of $\sim$ 53,000 galaxies at $z \sim$ 0.1 spanning over 3 orders of magnitude in stellar mass and a factor of 10 in metallicity \citep[see an update for star forming galaxies in][]{2016MNRAS.457.2929W}. The MZR has been established at different redshifts \citep[e.g.,][]{2005ApJ...635..260S, 2006ApJ...644..813E, 2008A&A...488..463M} and different environments \citep[e.g.,][]{2007MNRAS.382..801M, 2013A&A...550A.115H}. 

Several scenarios have been invoked to explain the MZR from variations in the initial mass function with mass \citep{2007MNRAS.375..673K}, metal-poor gas infall \citep{2008MNRAS.385.2181F}, outflows or accretion of gas \citep{2004ApJ...613..898T, 2005MNRAS.362...41G, 2007MNRAS.376.1465K} , selective increase of the star-formation efficiency with the stellar mass \citep[also known as \emph{downsizing},][]{2007ApJ...655L..17B, 2008ApJ...672L.107E, 2009A&A...504..373C, 2009MNRAS.396L..71V}, or a combination of these scenarios. 

The second key property is the existence of radial gradients in the gas-phase metallicity in star-forming galaxies. This too is a long-known and often-studied problem \citep[e.g.,][]{1994ApJ...420...87Z,2010ApJS..190..233M, 2012ApJ...745...66M, 2015MNRAS.454.3664B, 2015MNRAS.451..210C, 2014A&A...563A..49S,2016A&A...587A..70S}.

The MZR and radial metallicity gradients are usually analyzed separately, even though the physical and dynamical processes that have established them could be strongly related. In the present paper we are exploring the possibility that both the MZR and the radial metallicity gradients have a common origin that arises from a local empirical relationship between the metallicity and the local stellar surface mass density in the disk. This possibility is suggested by the strong positive correlation between the effective surface mass density and stellar mass \citep[e.g., ][]{2003MNRAS.341...33K} and a strong systematic decrease in the stellar surface mass density with increasing radius \citep[e.g., ][]{2015ApJ...800..120Z}.

The relationship between local metallicity and local surface mass density has been studied before. These studies reported that HII regions with larger stellar densities are more metal rich in comparison to those with lower densities \citep{1992MNRAS.259..121V,1984MNRAS.211..507E}. Later, \cite{2012ApJ...745...66M} showed a trend between the surface mass density and the metallicity in their sample of 174 star-forming galaxies using long-slit spectroscopy. This was further considered and verified by \cite{2015MNRAS.451..210C} for a sample of HI-rich galaxies. Recently, the integral-field spectroscopy (IFS) technique has become possible, allowing the analysis of the spatially resolved properties in relatively large samples of disk galaxies. \cite{2012ApJ...756L..31R} demonstrated the local existence of the surface-mass density vs. metallicity relation using IFS data from HII regions in 38 spiral galaxies. \cite{2013A&A...554A..58S} confirmed this local relation using a sample of 150 star forming galaxies included in the CALIFA survey \citep{2012A&A...538A...8S}. The above IFS studies also find a relation between the surface mass density and the specific star formation rate. 

Despite these efforts, the samples of these IFS surveys are rather small in size and cover a limited range of global galaxy properties. For instance, the stellar mass coverage of the CALIFA survey is complete for nearby galaxies with 9.5 < $\log(\mathrm{M}_*/M_{\odot})$ < 11.0 \citep{2014A&A...569A...1W}. A larger sample of galaxies covering a wider range of physical parameters will allow us to understand whether the relations derived locally are independent of physical parameters that can affect a galaxy as a whole. The MaNGA survey \citep[Mapping Nearby Galaxies at APO,][]{2015ApJ...798....7B} is ideal for probing the impact of global parameters on the local relations. This on-going survey aims to observe in the next 5 years around 10000 galaxies using the IFS technique. Currently more than 1300 galaxies have been observed, allowing us to explore spatially resolved information for a sample of disk galaxies that covers a wide range of stellar masses and star formation rates. 

By utilizing this large sample and the wide range it covers in galaxy properties, the present paper is aimed at testing a simple hypothesis: the local gas-phase metallicity at a given location in a disk galaxy is determined mainly by the local stellar surface mass density, irrespective of the mass or the star-formation rate of the galaxy in which it resides. If so, this would imply that at least some global scaling relations in galaxies can have a local origin.

This article is organized as follows. In Sec.~\ref{sec:SampleCubes} we describe the main properties of the sample of disk galaxies as well as the data used in this study. In Sec.~\ref{sec:Analisis} we present the derivation of the surface mass density, metallicity, and the selection of the star-forming spaxels to be analyzed. We will then present the \emph{local} surface mass density vs. metallicity relation for the MaNGA disk galaxies and compare it to previous IFS surveys. We will examine the residuals in this relation as a function of the total stellar mass and sSFR  (Secs.~\ref{sec:muZ_Mtot}) as well as the metallicity profiles derived from this relation ~\ref{sec:dOH_grad}). In Sec.~\ref{sec:Discussion} we discuss the results of this study. Our main results and conclusions are presented in Sec.\ref{sec:Summary}. We adopt $H_0$~=~70~km~s$^{-1}$~Mpc$^{-1}$,~$\Omega_M$~=~0.3 and $\Omega_\Lambda$~=~0.7.

\section{Sample and MaNGA data cubes}
\label{sec:SampleCubes}

\subsection{Sample Selection}
\label{sec:Sample}

The targets in this study have been selected from the galaxies observed in the MaNGA survey as of June 2015 (1350 unique galaxies). These galaxies are included in the fourth internal data release of the survey (MPL-4). The ongoing MaNGA survey \citep{2015ApJ...798....7B} has been designed to acquire integral-field spectroscopic information for 10,000 galaxies in the Local Universe. The main parameters for the selection are described in \cite{2015ApJ...798....7B}. A detailed description is presented in \cite{2016AAS...22733401W}. A large fraction of the MaNGA sample is drawn from the SDSS Main Galaxy Sample. The luminosity-dependent volume-limited MaNGA sample is defined by the following properties: (i) $\log( M_{*}/ M_{\odot}$)~>~9.0; (ii) similar number of galaxies at different stellar mass bins (i.e, a uniform coverage in the stellar mass regime); (iii) uniform radial coverage for the sample. The half-light radius ($\mathrm{R}_e$) of each galaxy is used to determine its extension. To accomplish different science goals, two subsamples have been defined. The Primary subsample is aimed to cover the region interior to 1.5 $\mathrm{R}_e$ in at least  80\% of this sample ($\sim$ 5000 galaxies with $z \sim$ 0.03). A Secondary Sample is designed to cover the region interior to 2.5 $\mathrm{R}_e$ ($\sim$ 3000 galaxies with $z \sim$ 0.045). The remaining 10\% of the total number of targets is devoted to ancillary observations. Estimates project that a total of $\sim$ 10,400 galaxies will be observed over the next 5 years. 

\begin{figure}
	 \includegraphics[width=\linewidth,natwidth=610,natheight=642]{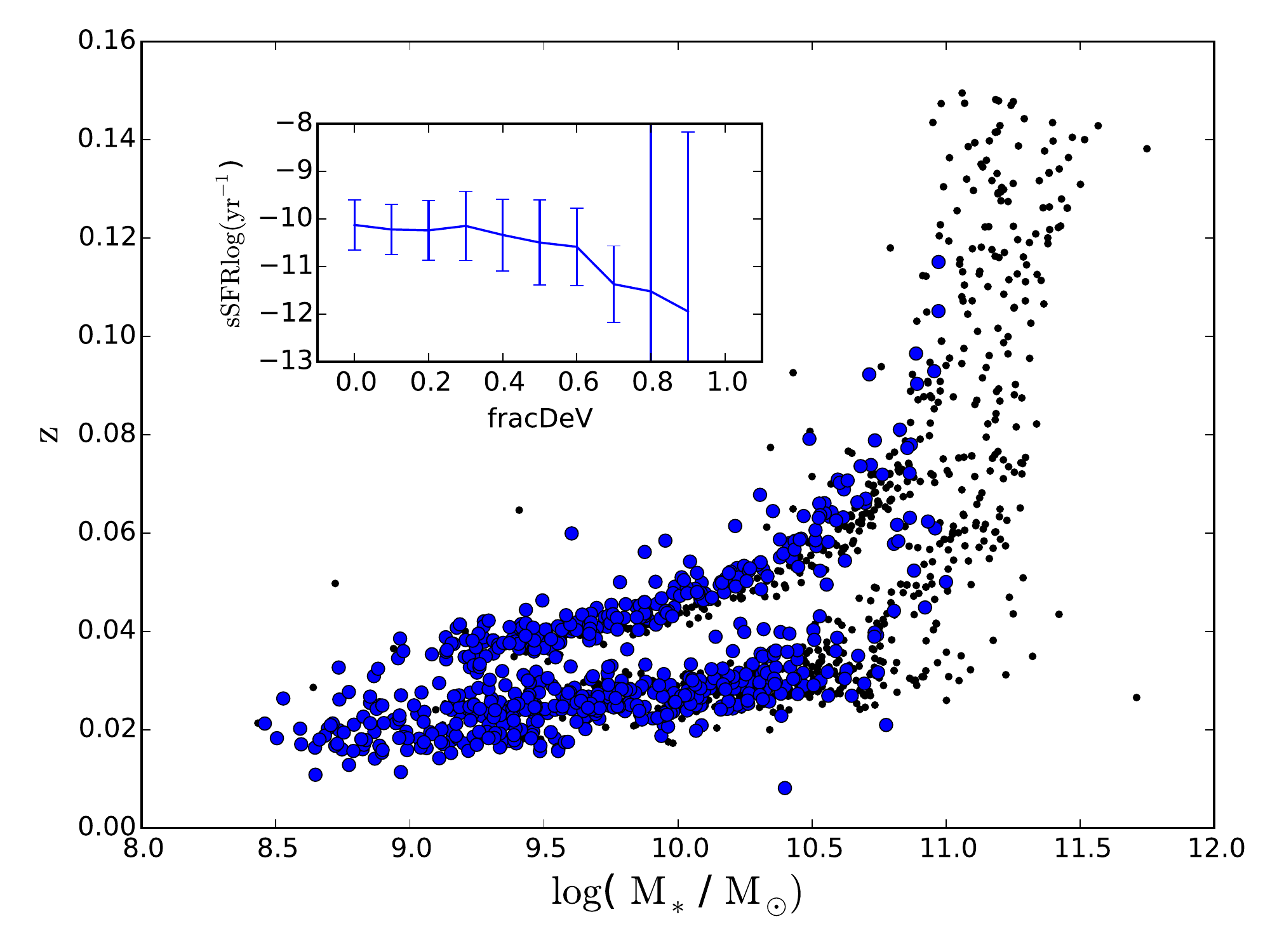}
    \caption{Distribution of our disks sample (blue circles) compared with the total sample of MaNGA observed galaxies (MPL-4, black dots) in the redshift-stellar mass plane. Note that due to our selection criteria (see details in Sec~\ref{sec:Sample}), our disk sample is biased towards lower mass/redshift values compared to the MPL-4 sample. The inset shows that total sSFR is similar for galaxies in the MPL-4 sample with \texttt{fracDeV} < 0.7 \citep[thereshold for disk galaxies, see]{2015ApJ...800..120Z}, at larger values the sSFR drops and the scatter increases.}
    \label{fig:Sample}
\end{figure}

We focus here on the analysis of spatially resolved properties in late-type galaxies in which the gas-phase metallicity can be mapped using emission-lines from star-forming regions. Accordingly, we select from the observed MaNGA sample by means of the morphology estimator \texttt{fracDeV}, defined in \cite{2005AJ....129.1755A}. This parameter quantifies if the light distribution of the galaxy is similar to either an exponential profile (low \texttt{fracDeV}) or to a de Vaucouleurs profile (high \texttt{fracDeV}). We select as disk galaxies those with a \texttt{fracDeV} < 0.7, following \cite{2015ApJ...800..120Z}. Note that similar values have been used to discriminate between spiral and elliptical in morphological studies \cite[e.g.,][]{2008MNRAS.388.1321P, 2013MNRAS.434.2153R}.

In Fig.~\ref{fig:Sample} we plot the stellar mass and redshift of galaxies from the MPL-4 sample. These two properties have been taken from the NSA catalog \footnote{NASA-Sloan Atlas (NSA; Blanton M. http://www.nsatlas.org). The NSA is a catalog of nearby galaxies within 200 Mpc (z = 0.055) based on the SDSS DR7 MAIN galaxy sample \citep{2009ApJS..182..543A}, but incorporating data from additional sources and image reprocessing optimized for large galaxies \citep{2011AJ....142...31B}. For MaNGA, the NSA has been extended to z = 0.15.}, from which the MaNGA main sample has been selected. Even though the distributions of stellar masses or redshift for the MPL-4 sample are not homogeneous, the sample does cover a  wider range of these two parameters than previous integral field surveys \citep[e.g. CALIFA, ][]{2014A&A...569A...1W}. Both distributions peak at $\log(M_{*}/ M_{\odot}) \sim 10.8$ and $z \sim$ 0.03, respectively. Our selection parameter for disk galaxies (\texttt{fracDeV} < 0.7) significantly reduces our sample, particularly at high stellar masses/redshifts. However it does guarantee that we are selecting galaxies consider that are star-forming (see inset in Fig.~\ref{fig:Sample}). To assure a significant number of galaxies in the different stellar mass bins we restrict our sample to galaxies with $\log(M_{*}/ M_{\odot})$~<~11.0 (see red-star simbols in Fig.~\ref{fig:Sample}). Using these criteria our sample of disk galaxies consists of 653 galaxies.

\subsection{MaNGA Cubes}
\label{sec:Cubes}

Observations from the ManGA survey have been performed using the SDSS 2.5-m telescope located at the Apache Point Observatory \citep{2006AJ....131.2332G}. For a detailed description of the MaNGA instrumentation the reader is referred to \cite{2015AJ....149...77D}. Here we highlight its main features. It consists of six cartridges, each with a plate of $\sim$ 3 degree diameter that can be mounted at the Cassegrain focus of the telescope. Each plate has been predrilled to plug a set of fiber bundles. Each fiber bundle corresponds to an integral field unit. These fibers are connected to a twin spectrograph with two arms (blue and red), allowing a large portion of the optical/near-IR wavelength regime (3600 \AA\, to 10300 \AA) to be observed simultaneously \citep{2013AJ....146...32S}. Its spectral resolution ranges from $R \sim$ 1600 to 2600 (for the blue and red arms, respectively).  Each plate has 17 science fiber bundles. For each of these bundles, the number of fibers ranges from 19 to 127. The diameter of each fiber is 2 arcsec. The fibers are arranged in each bundle in a hexagonal shape \citep[see Fig.7 in][]{2015AJ....149...77D}. To have a total coverage of the hexagonal field of view, a three point dithering pattern has been adopted, similar to other IFU surveys \citep[e.g., ][]{2012A&A...538A...8S}. An extra set of bundles is used to obtain the night-sky spectra. 

The MaNGA survey has been designed to ensure high quality and uniform coverage across the sample of observed galaxies \citep[for a detailed explanation see][]{2015AJ....150...19L}. To achieve this, the total exposure time for a given plate is typically 3 hours. Individual plates are observed until the combined depth reaches a S/N of 5 \AA$^{-1}$ fiber$^{-1}$ at a surface brightness of 23 AB arcsec$^{-2}$ in the SDSS r-band. These exposures are also required to be obtained in time windows for which the air mass is less than 1.21. Data reduction as well as observing strategy are described in detail in \citep{2016arXiv160708619L}. The MaNGA reduction pipeline includes wavelength calibration, corrections from fiber-to-fiber transmission, subtraction of the sky spectrum, and flux calibration \citep{2016AJ....151....8Y}. By combining the set of exposures in each pointing, the spectra are arranged in a datacube with the $x$ and $y$- axes corresponding to spatial location and the $z$- axis corresponds to wavelength. This final data cube has a spatial sampling of 0.5 arcsec/spaxel and a spatial resolution of 2.5 arcsec corresponding to mean physical scales of 1.5 and 2.5 kpc for the Primary or Secondary sample, respectively.  
\begin{figure}
    \includegraphics[width=\linewidth,natwidth=610,natheight=642]{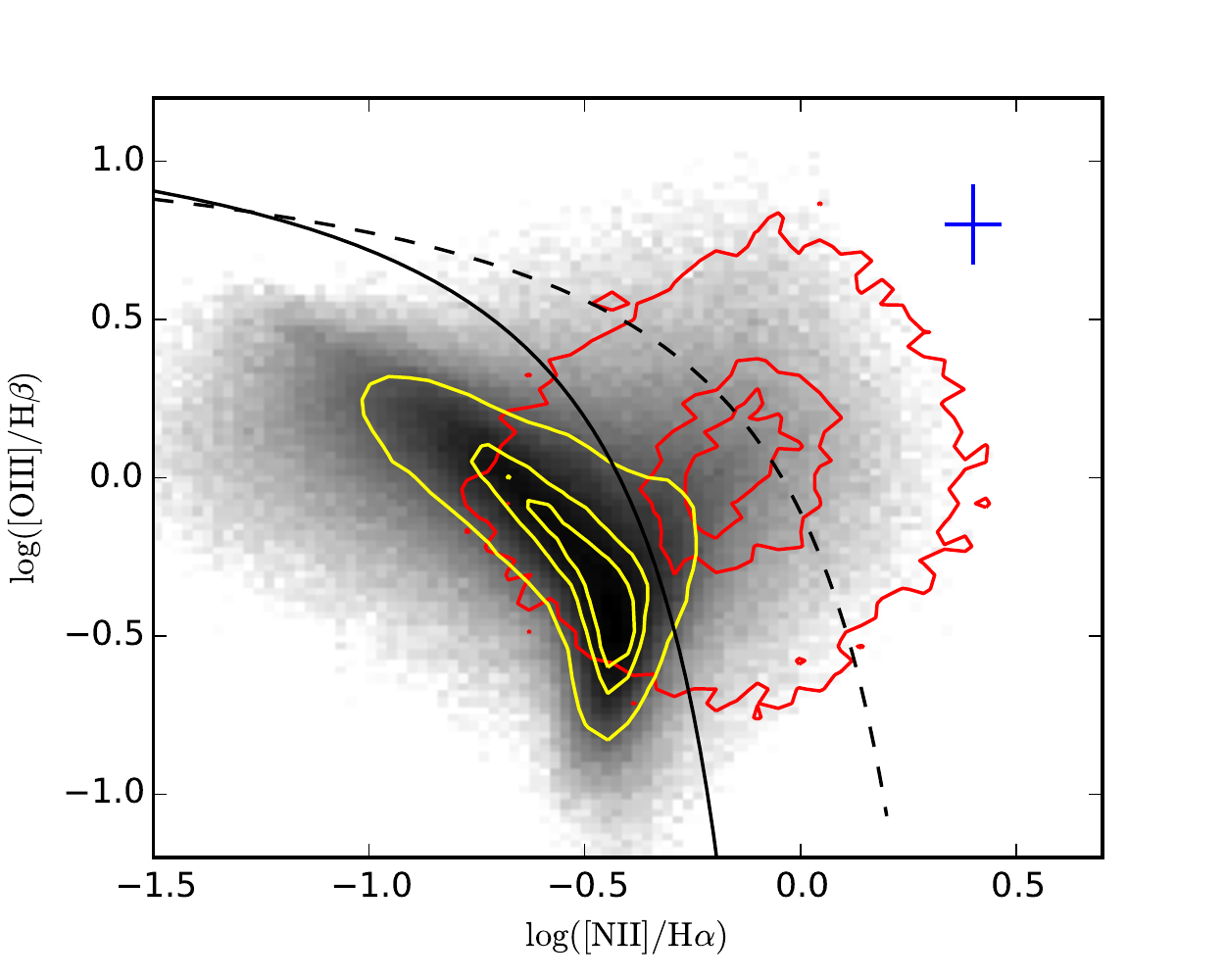}
   \caption{[OIII]/ H$\beta$ and [NII]/ H$\alpha$ diagnositic diagram for all the spaxel in our sample of disk galaxies. The yellow and red contours indicate those spaxels where their H$\alpha$ EW is larger and smaller than 6 \AA, respectively.}
    \label{fig:BPT}
\end{figure}

\section{Analysis and Results}
\label{sec:Analisis}
\subsection{Determination of local properties - PIPE3D}
\label{sec:PIPE3D}
\begin{figure*}
    \includegraphics[width=0.8\linewidth, natwidth=610,natheight=642]{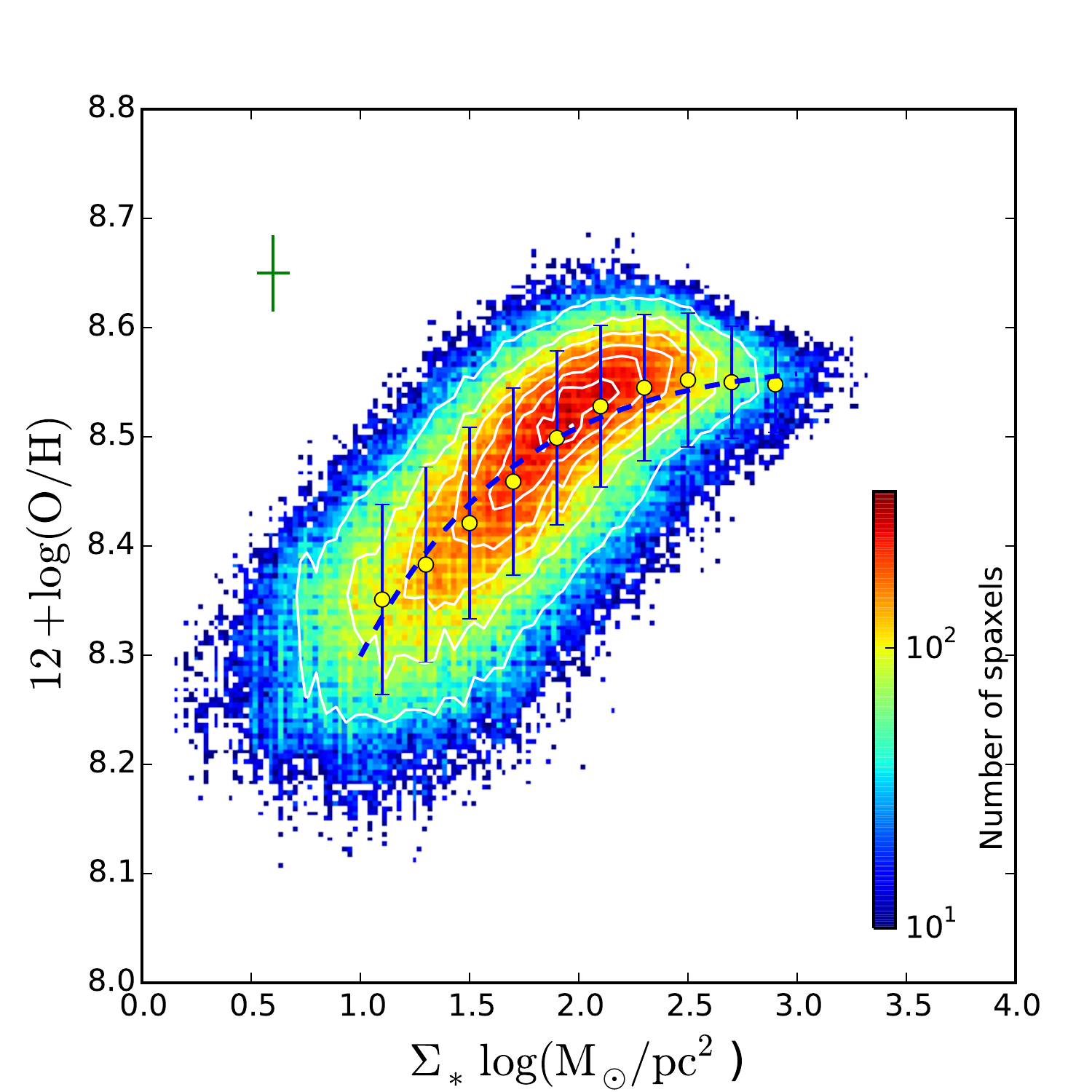}
    \caption{The distribution of the oxygen abundance for more than 507,000 star-forming spaxels against their stellar mass surface densities. These spaxels are extracted from 653 MaNGA galaxies considered as disks (see Sec.\ref{sec:Sample} for details). The colorbar shows the number of spaxels per bin in the $\Sigma_*$-Z space. The white outer contour encloses 80\% of the sample while each inner contours enclose $\sim$15\% less, consecutively. The yellow circles represent the median metallicity for spaxels with surface mass density within a bin width of 0.2 dex between 1.0 < $\Sigma_*$ < 3 dex. The blue error bars represent the standard deviation in metallicity for each of these bins. The dashed blue line represents the best-fitted curve for the yellow median circles.}
    \label{fig:mu-OH-ssfr}
\end{figure*}
With the observations reduced and the datacubes created, the next step consists of extracting the physical parameters from each of the spaxels in each galaxy in the form of two-dimension distributions. In particular, we are interested in the surface mass density and the emission line fluxes. To measure these, we use the PIPE3D pipeline \citep{2016arXiv160201830S}. This pipeline has been intended to be used for different scientific goals \citep{2013A&A...554A..58S, 2014A&A...563A..49S, 2016arXiv160202770C}. For a detailed description on the fitting procedure, dust attenuation and uncertainties determination see \citep{2015arXiv150908552S,2016arXiv160201830S}. Here we describe the main features of this analysis. 

As a first step, the cube is spatially binned to reach a signal-to-noise ratio (S/N) of 50 in the stellar continuum. The binning algorithm used in PIPE3D is based on both a continuity criterion in the surface brightness and a goal in the S/N. The stellar properties in each bin are derived by fitting a set of single stellar population (SSP) templates to the stellar continuum. For the MaNGA data cubes the pipeline uses a grid of templates extracted from the MIUSCAT SSP library \citep{2012MNRAS.424..157V}. The grid covers four stellar ages (0.06, 0.20, 2.00,and 17.78 Gyr) and three metallicities (0.0004, 0.02, 0.0331). Specifics in the fitting procedure including foreground dust attenuation, and uncertainties are described in \citep{2016arXiv160201830S}. The stellar population model for each spaxel is obtained by rescaling the best fitted model in each bin to the continuum flux intensity corresponding to each spaxel. The stellar mass is obtained by assuming the same mass to light ratio and the same attenuation for the spaxels included in each bin. The typical error for the estimated stellar mass in each spaxel is smaller than 0.15 dex. The spaxel surface mass density ($\Sigma_*$) is thus obtained using the ratio of the stellar mass within the spaxel and its deprojected area. To correct each spaxel's surface mass density from projection effects, we multiply the observed density by a factor $b/a$, where $a$ and $b$ represent the projected semi-major and semi-minor axis of the disk galaxy, respectively. We obtain these photometric parameters from the NSA catalog. This correction factor comes from assuming an axisymmetric mass distribution (e.g., thick disk or an oblate spheroidal), the observed total area is $A_\mathrm{obs} = \pi a b$, while the total area of the disk (as seen face on) is $A = \pi a^2$. In average, the surface mass density is given by:
\begin{equation}
\Sigma_* = \frac{M}{\pi a^2} = \frac{M}{A_\mathrm{obs}} = \Sigma_{*,obs} b/a 
\end{equation}
where $M$ is the total mass of the disk and $\Sigma_{*,obs}$ is the observed surface mass density. In the next section we show that the scatter of the local mass metallicity relation is reduced by considering the deprojected area for each spaxel.

The ionized gas data-cube is created by subtracting the stellar population model in each spaxel. The strongest emission lines within the MaNGA wavelength regime are fitted on a spaxel-by-spaxel basis. The pipeline provides 2-D flux, intensity, and equivalent width (EW) maps of a large set of emission lines including H$\alpha$, H$\beta$, [OIII]$\lambda$5007, and [NII]$\lambda$6548. We use the flux ratio of these emission line to determine the oxygen abundance of the star forming spaxels in our sample of disk galaxies. 

We select the star forming spaxels in each of the galaxies by comparing their [OIII]/H$\beta$ and [NII]/H$\alpha$ line ratios in a BPT diagnostic diagram \citep[][see Fig.~\ref{fig:BPT}]{1981PASP...93....5B}.  Following \cite{2014A&A...563A..49S}, we classify the star forming spaxels as those with line ratios below the Kewley demarcation curve \citep[dotted curve in Fig~\ref{fig:BPT} ][]{2001ApJ...556..121K} and H$\alpha$ EW larger than 6 \AA.\, We note that the combination of these two criteria is a reliable method to select star forming spaxels. Even more, a large fraction of spaxels with EW larger than 6 \AA\, are already below the Kauffmann demarcation curve \citep[solid curve in Fig~\ref{fig:BPT}][]{2003MNRAS.346.1055K}. Most of the spaxels in our sample of disk galaxies have EW's larger than 6 \AA\, ($\sim$ 86\%). 

Once we selected our sample of star forming spaxels, we can use the metallicity calibrators derived from HII regions studies. For this particular study, we use the metallicity calibrator from \cite{2013A&A...559A.114M} They determine the metallicity  using the $O3N2$ calibrator. This calibrator uses the logarithmic difference between the line ratios log([OIII]/H$\beta$) and log([NII]/H$\alpha$) :
\begin{equation}
12 + \log\mathrm{(O/H)} = 8.533[\pm 0.012] - 0.214[\pm 0.012]\times O3N2 
\end{equation}
Typical errors for metallicity from the flux ratio calibrator is $\sim$ 0.06 dex.

\subsection{The  MaNGA Surface Mass Density - Metallicity ($\Sigma_*$-Z) relation}
\label{sec:localMZ}
In Fig.~\ref{fig:mu-OH-ssfr} we plot the distribution of the oxygen abundance for the star forming spaxels as function of their surface mass density for our sample of MaNGA disk galaxies. This distribution includes more than 507,000 star forming spaxels. To our knowledge, this is the largest sample of spatially resolved elements for disk galaxies to date. The oxygen abundance shows a clear correlation with the surface mass density that extends over three orders of magnitude in the latter and a factor of four in metallicity. Higher surface mass densities exhibit progressively higher metallicity. It is remarkable how these two {\it local} properties correlate in a similar manner as their global counterparts, as reported in the literature \citep[e.g.,][ see also Sec~\ref{sec:totalMZ}]{2004ApJ...613..898T,2012A&A...547A..79F,2016MNRAS.457.2929W}. This relation is tight: metallicity has a scatter smaller than 0.08 dex with respect to its median value at different mass density bins (see error bars in Fig.~\ref{fig:mu-OH-ssfr}). We perform a fitting to the median abundance at different mass densities. Following \cite{2013A&A...554A..58S}, we use the following fitting function:
\begin{equation}
y = a + b (x - c) \, e^{-(x-c)}
\end{equation}
where $y$ = 12+$\log$(O/H) and $x$ is the logarithm of the stellar mass density ($\Sigma_*$) in units of $M_{\odot}$/$\mathrm{pc}^{2}$. This functional form has been selected in order to describe the observed distribution of the metallicity with respect to the mass density, which seems to show a saturation in metallicities at the highest densities. The coefficients corresponding to the best fit of the above function to the MaNGA galaxies presented in this study are $a$ = 8.55 $\pm$ 0.02, $b$ = 0.014 $\pm$ 0.02 and $c$ = 3.14 $\pm$ 1.87.   
\begin{figure}
    \includegraphics[width=\linewidth,natwidth=610,natheight=642]{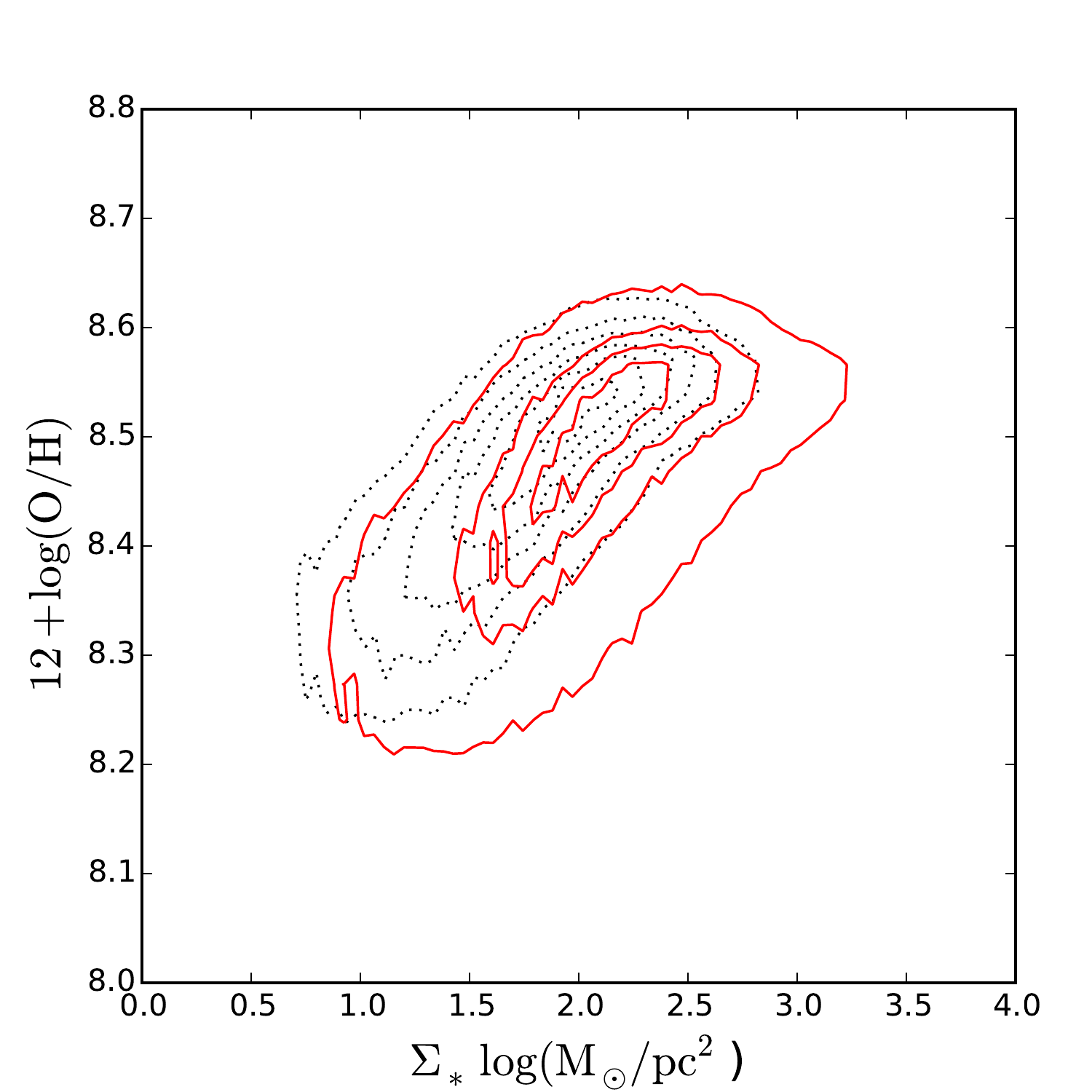}
    \caption{The $\Sigma_*$-Z relation using the inclination corrected $\Sigma$ (black dotted contours) and the observed surface mass density ($\Sigma_{*,obs}$, red solid contours). It is evident that the scatter of the relation is larger when using the observed surface mass density.}
    \label{fig:Sigma-OH-corrected}
\end{figure}

 In Fig~\ref{fig:Sigma-OH-corrected} we compare the relation in Fig.~\ref{fig:mu-OH-ssfr} (dotted black contours) with the relation derived using $\Sigma_{*,obs}$ (red solid contours). The scatter of the $\Sigma_{*,obs}-Z$ relation is considerable larger than the one presented previously. Since we use the residuals of this relation to investigate possible dependence of the global parameters of disk galaxies in the metal content of spatially resolved regions (see Sec.\ref{sec:muZ_Mtot}), it is essential to reduce potential systematic effects  that could alter the metallicity residuals.
\begin{figure}
    \includegraphics[width=\linewidth]{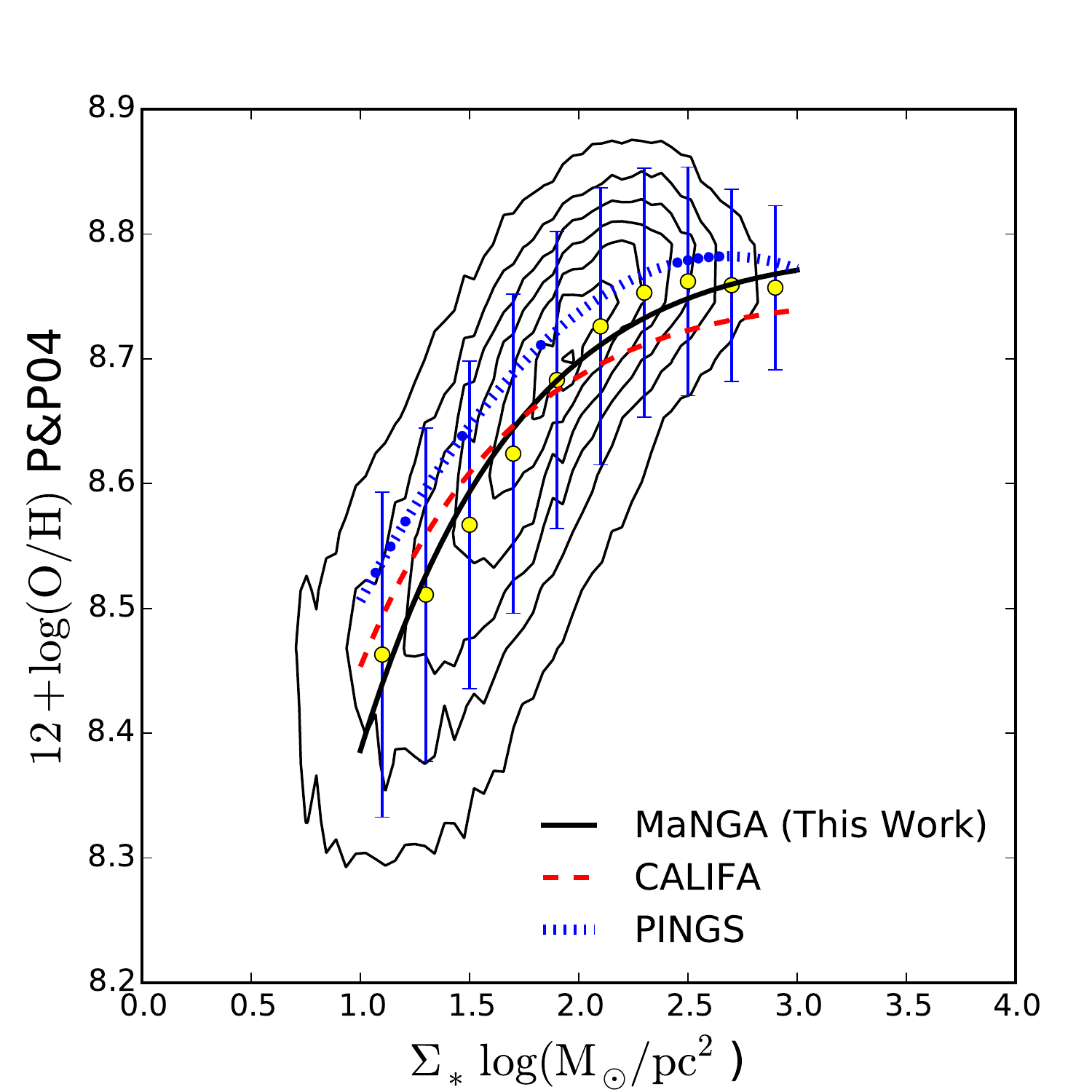}
    \caption{Comparison of the $\Sigma_*$-Z for different IFS surveys. In order to compare our measurements with those from previous IFS surveys we use the metalicity calibrator from Pettini \& Pegel (2004). Similar to the Fig.~\ref{fig:mu-OH-ssfr}, black contours enclose the same fraction of spaxels. Yellow circles represent the median metallicity for spaxels with surface mass density binned in 0.2 dex with the error bars showing their metallicity standard deviation.}
     
    \label{fig:muZ-Surveys}
\end{figure}

The relation between the local surface mass density and metallicity described here is similar to the one reported from individual HII regions in previous IFS surveys. In Fig.~\ref{fig:muZ-Surveys} we compare the distribution of the $\Sigma_*$-Z relation presented in this study and the best fitted-curves from the PINGS \citep{2012ApJ...756L..31R} and the CALIFA surveys \citep[][]{2013A&A...554A..58S} with red-dashed and blue-dotted lines, respectively. In order to make a similar comparison between the relation presented here and those in previous IFS surveys we use the oxygen abundance calibrator in \cite{2004MNRAS.348L..59P}. Within the scatter of the relation, the shape of the $\Sigma_*$-Z relation for the different surveys are quite similar. The fitting from the PINGS survey shows a positive offset of $\sim$ 0.11 dex in comparison with the MaNGA best fitted curve. This offset is very small at large mass densities. The CALIFA fit yields slightly larger (smaller) metallicities at low (high) mass densities in comparison with the curve presented in this study (by +0.06 and -0.03 dex, respectively). Note that our relationship is based on a much larger sample than these previous studies yet the local distributions are similar: whereas the PINGS and CALIFA surveys included 38 and 150 disk galaxies respectively, we include 653 galaxies from the MaNGA sample.

\subsection{The MZR from MaNGA Datacubes}
\label{sec:totalMZ}
\begin{figure}
    \includegraphics[width=\linewidth,natwidth=610,natheight=642]{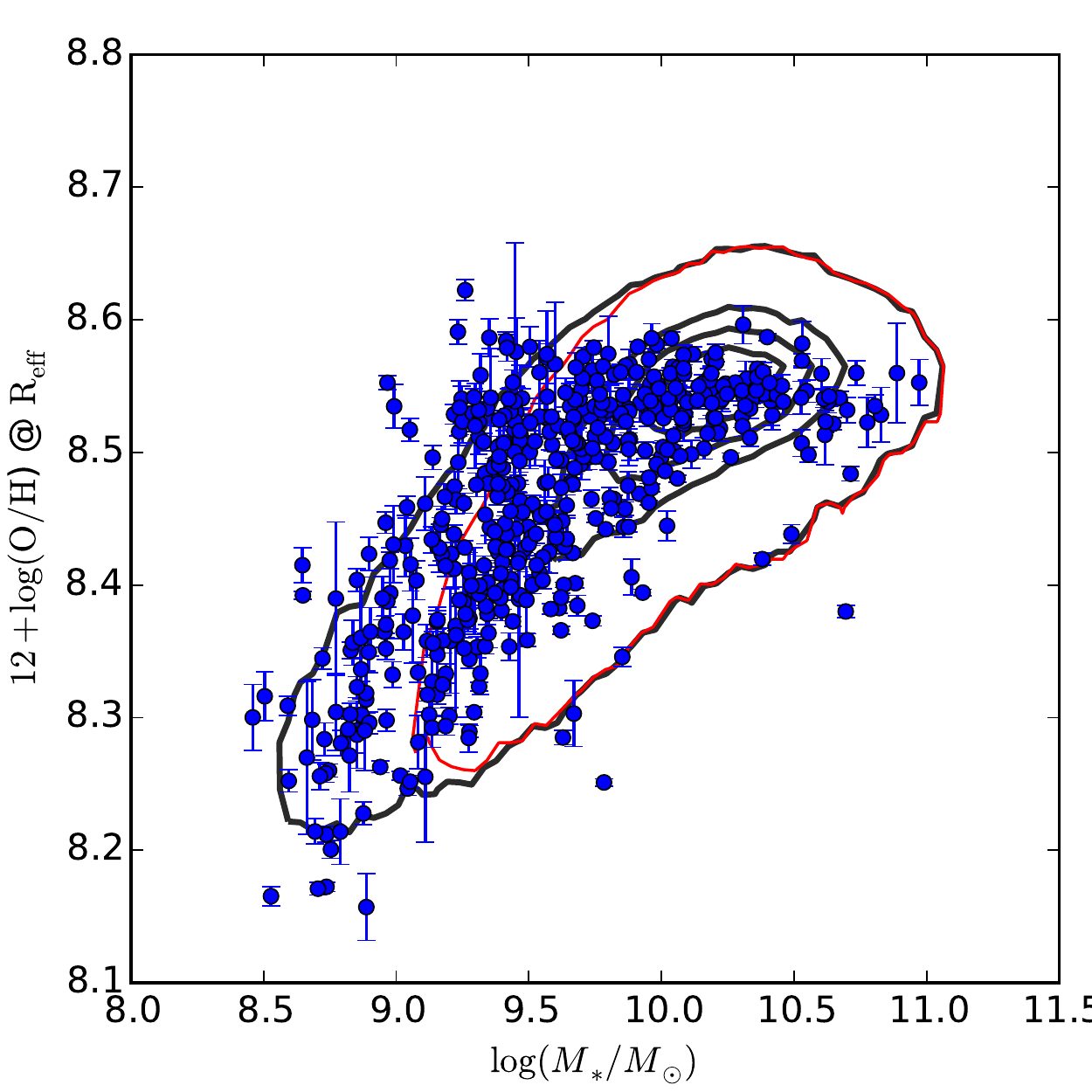}
    \caption{The MZR for MaNGA galaxies. The metallicity for each galaxy has been derived using the integrated values within an effective radius from the two-dimensional distributions of the emission line fluxes. Stellar masses are drawn from the NSA catalog. Black contours show the MZR distribution from  SDSS galaxies in the same stellar mass and redshift ranges as those cover by our MaNGA sample. For comparison, the red and gray contours show the MZR distribution at the redshift range presented by \citep{2010A&A...521L..53L} and \citep{2010MNRAS.408.2115M}. Contours cover 95 \% of the SDSS sample.   }
    \label{fig:TotalMZR}
\end{figure}
We want to compare how the global properties derived by integrating over the two-dimentional distributions of our sample compare to those derived by single-fiber spectroscopy. For this exercise in each galaxy of our sample we integrated within an effective radius ($\mathrm{R_{eff}}$) the total flux in each of the two-dimensional distribution corresponding to the emission lines required to determine our metallicity indicator (see Sec.~\ref{sec:PIPE3D}). We select the galaxies considered as star forming from a BPT diagnostic diagram using the ratios from the integrated fluxes. From this selection we derived the metallicity in 480 galaxies. The uncertainties are also derived using the error maps from each emission line. We do not consider the systematic error from the abundance calibrator.  

In Fig.~\ref{fig:TotalMZR} we plot the MZR for the MaNGA galaxies. For comparison purposes only, we over plot the contours of the MZR distribution derived from single-fiber spectroscopy in the galaxies included in the SDSS sample. Black contours cover the same redshift and stellar mass ranges as our sample while the red contour represents the MZR in a study covering similar redshift range  as our sample \citep[][, see also \cite{2012A&A...547A..79F}]{2010A&A...521L..53L}. For this sample we derived the metallicity using the fluxes and stellar masses from the Max-Planck-Institute-Johns-Hopkins-University (MPA-JHU) emission-line analysis database \footnote{\texttt{http://www.mpa-garching.mpg.de/SDSS/DR7}}. The MaNGA integrated metallicities cover similar values to those covered by single-fiber spectroscopy within the same stellar mass / redshift regime. In comparison with previous studies, the MaNGA sample cover galaxies at lower redshift, as a result, both the MaNGA data points and black contours cover low-mass galaxies. Despite the difference between methods, covered area, and ranges of the different samples the MZR constructed from integrated fluxes from the MaNGA spatially resolved data follows quite a similar trend as those reported in the literature derived from single fiber spectroscopy. We note a small group of outliers with an unusual large metallicity for a low-mass galaxy (~$10^{9.2} \mathrm{M}_\odot$). Most of these targets are either compact galaxies or disk with high inclination. They also show a evident bulge. Since these galaxies are compact, we suspect that the integrated metallicity at 1 R$_\mathrm{eff}$ includes a wider portion of the galaxy than their conterparts at similar stellar mass range.  

\begin{figure*}
     \includegraphics[width=\linewidth,natwidth=610,natheight=642]{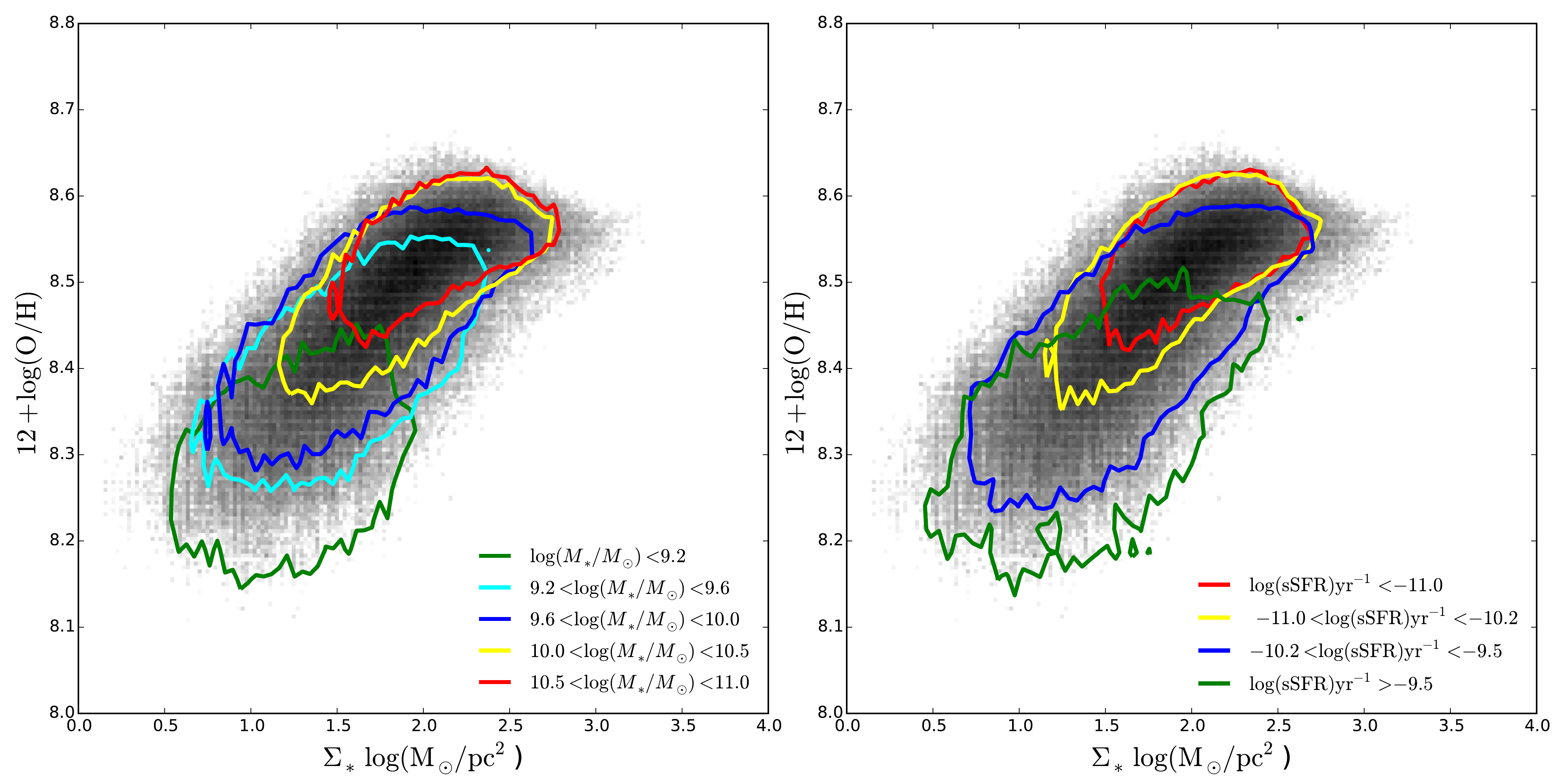}
    \caption{$\Sigma$-Z relation distribution divided at different total stellar mass and total specific SFR (sSFR) bins (left and right panels, respectively). In each of the stellar mass/sSFR bins the contours encloses 80\% of the sample. Except for the lowest mass bin, the size of each stellar mass bin is $\sim$ 0.4 dex, ranging from 9.2 to 11.0. The sSFR is divided in four different bins from low ($\log(sSFR)$ < -11 $\log(\mathrm{yr}^{-1})$) to high ($\log(sSFR)$ > -9.5 $\log(\mathrm{yr}^{-1})$) values. For reference, we plot in gray colorscale the distribution presented in Fig.\ref{fig:mu-OH-ssfr}.}
    \label{fig:MuOH_conto}
\end{figure*}
\subsection{Residuals in the $\Sigma_*$-Z Relation}
\label{sec:muZ_Mtot}
While the $\Sigma_*$-Z relation has been described before, the much larger amount of data from MaNGA allows us to test the degree to which this relation -- on its own -- can adequately explain {\it both} the mass-metallicity relation and the radial metallicity gradients in disk galaxies.

\begin{figure}
     \includegraphics[width=\linewidth,natwidth=610,natheight=642]{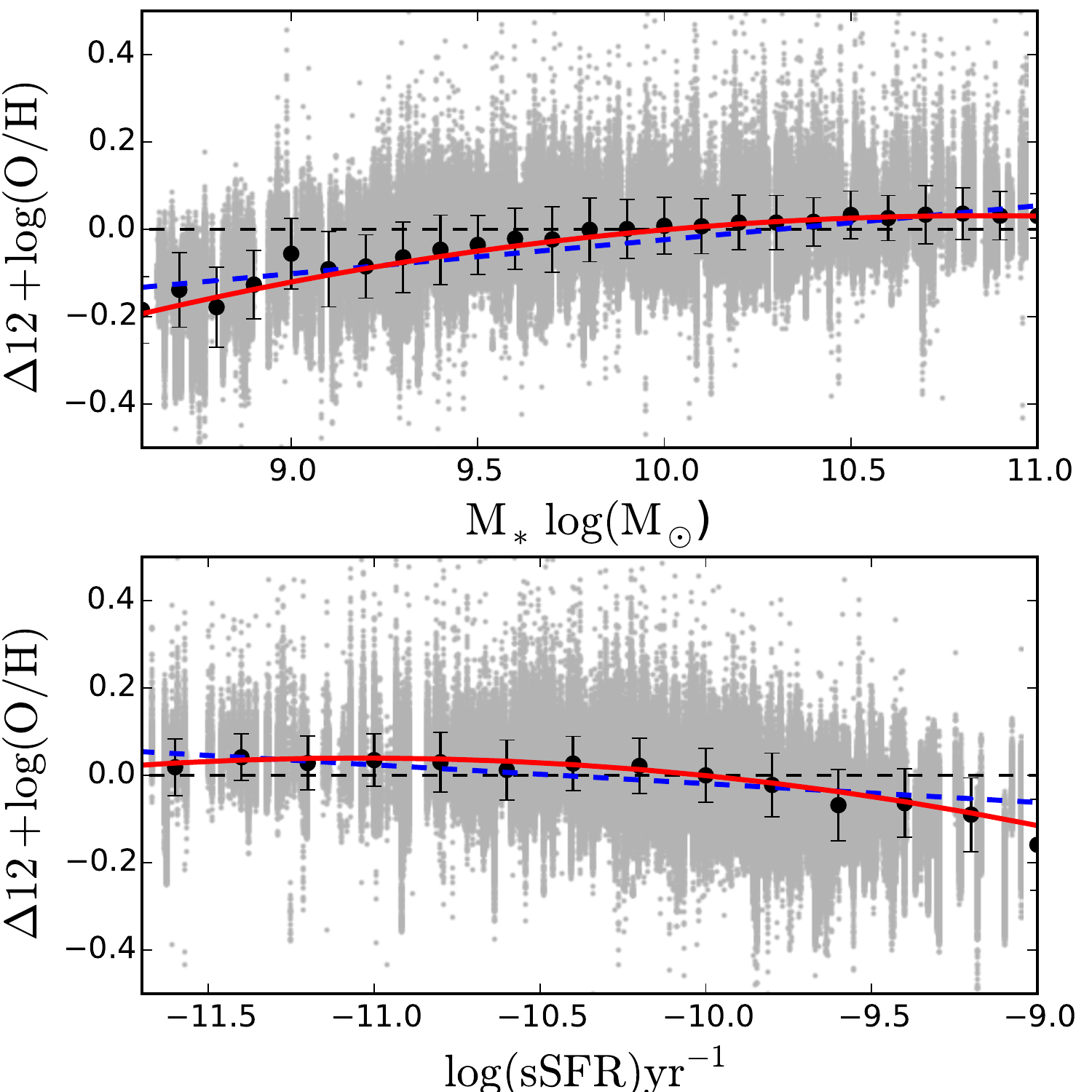}
    \caption{Scatter of the $\Sigma_*$-Z relation with respect to the total stellar mass. Black points represent the median value at different bins (0.1 dex for the total stellar mass). In both panels, the solid-red and dashed-blue curves represent a quadratic and linear fit of the median scatter metallicities, respectively. Black dashed line represents zero-scatter in the $\Sigma_*$-Z relation.}
    \label{fig:difOH_sfr}
\end{figure}

To start, we can ask how galaxies with different stellar masses populate the $\Sigma_*$-Z relation. To do this, we separated our sample in five bins with approximately the same width in total stellar mass distribution ($\sim$ 0.4 dex). In the left panel of Fig.~\ref{fig:MuOH_conto} we plot the $\Sigma_*$-Z relation for each of these stellar mass bins. This shows that lower-mass galaxies are composed of lower-metallicity material residing in regions of lower surface mass density. As the stellar mass increases the distribution of points in the $\Sigma_*$-Z relation move steadily along the relation towards progressively higher metallicities and surface-mass densities.  This shows explicitly how the relations between $\Sigma_*$, Z, and stellar mass can arise from purely local relations. 

Another important property of a galaxy is its stellar age. We parameterize this by the specific star formation rate (sSFR: the SFR per unit stellar mass). In the right panel of Fig.~\ref{fig:MuOH_conto} we divide our MaNGA sample into four bins in the sSFR measured for the galaxy based on the SDSS MPA-JHU catalog. As in the left panel, there is a smooth trend with galaxies having progressively lower values of sSFR as $\Sigma_*$ and Z increase. 

\begin{figure*}
     \includegraphics[width=\linewidth,natwidth=610,natheight=642]{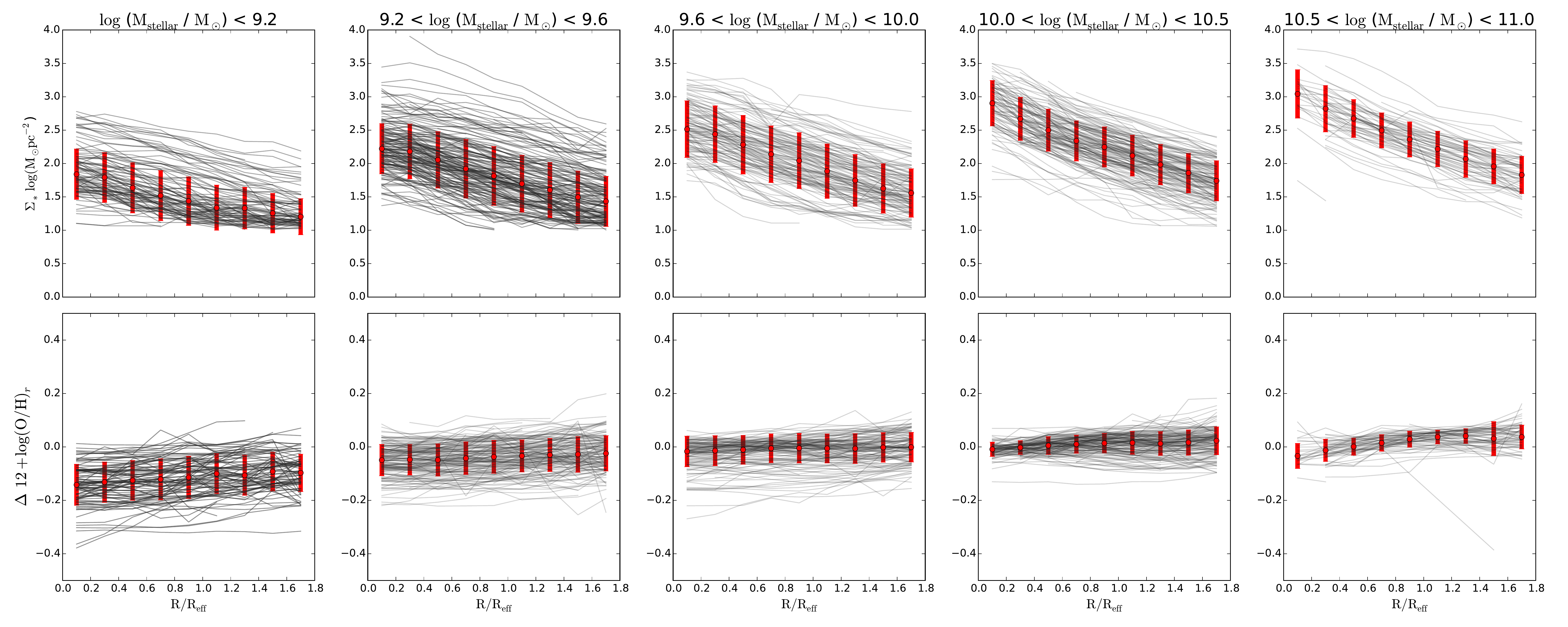}
    \caption{Top: Radial distribution of the surface mass density for disk galaxies at different total stellar mass bins in effective radius units. The mass bins are the same used in Sec.~\ref{sec:muZ_Mtot}. Gray lines represent individual gradients in radial bins of 0.2 $R_{\mathrm{eff}}$. For each radial bin red point and error bar represent the median and standard deviation, respectively. Bottom: Radial distribution of the difference between the observed metallicity and the value derived from the best fitted curve of the $\Sigma_*$-Z relation for the same mass bins within the range of $\Sigma_*$ where the fitting was performed. Red points and error bars represent the median and standard deviation for a given radius. Except for the lowest mass bin, the derived metallicity gradient is consistent to the observed one.}
    \label{fig:difOH_grad}
\end{figure*}

We now want to test explicitly whether the residuals in the best-fit $\Sigma_*$-Z relation correlate systematically with other properties of the galaxy. For each star forming region we calculate the difference between its observed metallicity and the one derived using the best fitted curve for the $\Sigma_*$-Z relation ($\Delta \, 12+\log(\mathrm{O/H})$). 

In top panel of Fig.~\ref{fig:difOH_sfr} we plot this scatter against the total stellar mass of the galaxy. We find that the residuals in this relation are small, and only exceed the scatter in the $\Sigma_*$ - Z relation (0.08 dex) for the lowest mass galaxies ($M_* < 10^{9.2} M_{\odot}$). A quadratic fit is better than a linear regression (red-solid vs dashed-blue lines). In a future paper we will further explore the physical origin of the effect at low stellar masses.

Following the same procedure outlined above for the stellar mass, we study how the scatter of the relation changes for different values of the global specific SFR computed from the MaNGA data. Again, the residuals are small, and only exceed the scatter in the $\Sigma_*$ - Z relation for the highest specific star-formation rates ($\mathrm{sSFR} > 10^{-9.2}$ yr$^{-1}$, corresponding to the starburst regime).

In conclusion, our results suggest that the gas-phase metallicity of star forming regions in disk galaxies is determined {\it primarily} by the local stellar surface mass density, rather than by other properties of the galaxies such as the total stellar mass or specific SFR.

\subsection{Radial Metallicity Gradients and the $\Sigma_*$-Z Relation}
\label{sec:dOH_grad}
%
In addition to the tight relation between the local stellar surface mass density and gas-phase metallicity ($\Sigma_*-Z$), disk galaxies also show tight relations between their total stellar mass and their effective stellar surface mass density ($\Sigma_{*,\mathrm{eff}} = M_*/2 \pi R_{\mathrm{eff}}^2$), and between their total stellar mass and their gas-phase metallicity (MZR). One may think that the correlation between these two \textit{local} parameters ($\Sigma_*$ and metallicity) could be a secondary correlation that is induced by these two other \textit{global} correlations. In this section, we will argue that this is not the case, because we will show that the $\Sigma_*-Z$ relation can also reproduce the well-known radial metallicity gradients seen in disk galaxies, which is independent of the MZR and $\Sigma_{*,eff}$ relations. To do so, we will use the spatially-resolved information from the MaNGA survey to determine how the observed radial metallicity gradients naturally emerge from the radial gradients in $\Sigma_*$ and the local $\Sigma_*-Z$ relation.

In top panels of Fig.~\ref{fig:difOH_grad} we show the radial profiles of the surface mass density in the five different stellar mass bins introduced in Sec.~\ref{sec:muZ_Mtot}. These profiles are obtained for each galaxy by averaging the surface mass density from the star forming regions within elliptical rings of width 0.2 $\mathrm{R_{eff}}$. These rings cover the radial range from 0.2 to 1.8 $\mathrm{R_{eff}}$ with the ellipticity and position angle drawn from the NSA catalog. We select this range of radii to cover a similar portion in each galaxy regardless of their selection criteria (Primary or Secondary, see Sec.~\ref{sec:Sample}). 

We then use the observed radial profile in $\Sigma_*$ and our best-fit $\Sigma_*$-Z relation to predict a radial metallicity gradient. In each of these rings we also measure the average metallicity of the star forming regions. In the bottom panels of Fig.~\ref{fig:difOH_grad} we plot the radial profiles of the differences between the measured and predicted metallicities ($\Delta \, 12+\log(\mathrm{O/H})_{r}$). We find that the observed and predicted metallicities are in good agreement over the range of radial locations and total stellar masses that we probe. In particular for the higher mass bins (three bottom rightmost panels in Fig.~\ref{fig:difOH_grad}), all median differences at different radii are smaller than the nominal error in the determination of the metallicity from emission line ratios ($\sim$ 0.06 dex), and the standard deviations of the residuals are of the same order ($\sim$ 0.04 dex). For intermediate masses (from 9.2 to 10.0 $\log(\mathrm{M}_*/\mathrm{M}_{\odot}$) the median differences are also quite small (~ 0.02 dex), however with larger standard deviations ($\sim$ 0.07 dex). For the lowest mass bin galaxies (bottom leftmost panel in Fig.~\ref{fig:difOH_grad}), the central observed metallicity is lower than the one derived from the central mass density  ($\sim$ 0.1 dex). This difference gets reduced at larger radii. The standard deviation is in general large compared to other mass bins ($\sim$ 0.1 dex). This consistent with the results in the top panel of Fig~\ref{fig:difOH_sfr}.

In conclusion, our study of the difference between the observed metallicity and those derived from the $\Sigma_*$-Z relation shows that for a wide range of stellar masses, the metallicity gradient observed in disk galaxies can be well reproduced by means of the radial density gradient taken together with the $\Sigma_*$-Z relation.

\section{Discussion}
\label{sec:Discussion}

We have confirmed that for our sample of disk galaxies the local oxygen abundance has a tight correlation with the local stellar surface mass density (see Fig.~\ref{fig:mu-OH-ssfr}). This is in agreement with previous IFS surveys \citep[see Fig.~\ref{fig:muZ-Surveys}, ][]{2012ApJ...756L..31R,2013A&A...554A..58S}.  Our results confirm that the well-studied MZR is a scaled-up version of the local $\Sigma_*$-Z relation. We have shown that the $\Sigma_*$ vs. Z relation can also explain the observed radial metallicity gradients. These results suggest that both the MZR relation and radial gradients are reflections of a more fundamental and intimate relation between local properties within disk galaxies. 

The much larger body of data provided by the MaNGA survey compared to prior IFS surveys makes it suitable for further analysis of the impact of global properties of the galaxy on the metallicity derived at local scales. Our results indicate that local oxygen abundance can be described by means of the best-fit of the $\Sigma_*$-Z relation over wide ranges of stellar mass and sSFR (covering almost two orders of magnitude in each parameter, see Fig~\ref{fig:difOH_sfr}). These results indicate that to first order the metal content of a disk galaxy depends more strongly on the local properties than on those of the galaxy as a whole. Similarly, in Sec.~\ref{sec:dOH_grad} we analyzed the differences between the observed metallicity gradients and those derived using the radial surface mass density gradients and the $\Sigma_*$-Z relation for galaxies in different stellar mass bins. Except for the lowest-mass galaxies ($\log(\mathrm{M}_*/M_{\odot})$ < 9.2), the derived radial metallicity gradients are in good agreement (see Fig.~\ref{fig:difOH_grad}). 

An interplay of different processes has been proposed to explain the chemical evolution of galaxies, including both the MZR and radial metallicity gradients \citep[e.g.,][]{2012MNRAS.421...98D,2013ApJ...772..119L}. The simplest case is the closed-box model in which the metal content of disk galaxies depends only on the metal yield ($y_Z$) and the gas mass fraction ($f_{\mathrm{gas}} = M_{\mathrm{gas}}/[M_* + M_{\mathrm{gas}}]$): $Z = y_Z\, \ln (f_{\mathrm{gas}})$. There is a strong observed inverse correlation between the local values for $f_{\mathrm{gas}}$ and $\Sigma_*$, and this then implies an observed strong inverse correlation between $f_{\mathrm{gas}}$ and $Z$ \citep{2015MNRAS.451..210C}. Thus, part of the $\Sigma_*$ vs. Z relation can be explained by the fact that the lower density regions are more gas rich and therefore simply less chemically-evolved. This is consistent with the standard ``inside-out'' model for the formation of disk galaxies in which the inner (denser) regions form earlier than the outer (less dense) regions which in turn are more chemically-evolved and less gas-rich. It is also consistent with the downsizing phenomenon in which the (progenitors of) more (less) massive galaxies form earlier (later).  

According to integrated observations, the metallicity should also be regulated by the gas accretion of metal-poor gas and/or ejection of metal rich material \citep[e.g., ][]{2004ApJ...613..898T,2005MNRAS.362...41G,2007MNRAS.376.1465K,2010MNRAS.408.2115M,2010A&A...521L..53L}. Indeed, we observed that the residuals in the $\Sigma_*$ vs. Z  at the lowest stellar masses (and highest values of sSFR) may reflect the effect of outflows, given that such galaxies have shallow potential wells and may therefore be less able to retain the metals they produce \citep[e.g.,][]{1974MNRAS.169..229L}.

These ideas have recently been considered in the context of two related scenarios for the origin of radial metallicity gradients. \cite{2015MNRAS.451..210C} proposed a scaled-down radial version of the reservoir model \citep[e.g., ][]{2013ApJ...772..119L} in which the metal content in each radial bin of a disk galaxy is regulated by local inflows of metal-poor gas from the halo, the metal rich material carried out of the galaxy by feedback, and the metals produced by massive stars (the stellar yield). With this simple model they were able to fit with relatively good agreement the radial metallicity profiles of 50 nearby galaxies with different contents of molecular gas \citep{2014MNRAS.441.2159W}. Similarly, \cite{2015MNRAS.448.2030H} suggest that metallicity gradients are the result of the coevolution of the stellar and gas disk in a virtual closed-box chemical evolution model. 

In our next paper, we will explore in more detail the inter-dependences between all the basic {\it local} properties of disk galaxies: the metallicity, the stellar surface-mass density, the escape velocity, the specific SFR, and the luminosity-weighted mean age. This in turn will allow us to gain a better understanding of the physical origin of the global mass-metallicity relation and radial metallicity gradients, and of the implications for galaxy evolution.

\section{Summary}
\label{sec:Summary}

We have used data from the MaNGA survey to study the role of the {\it local} stellar surface mass density $\Sigma_*$ in determining both the {\it local}metallicity relation and radial metallicity gradients in disk galaxies. While the relationship between $\Sigma_*$ and metallicity (Z) has been known for some time \citep[e.g.,][]{2012ApJ...745...66M, 2012ApJ...756L..31R, 2013A&A...554A..58S}, the much larger data set provided by MaNGA allows us to assess whether this relationship is a universal one, or whether instead the residuals in the relationship correlate systematically with the global properties of galaxies.
We found:
\begin{itemize}
\item
In agreement with these earlier studies, the metallicity increases steeply with increasing surface mass density over the range $\Sigma_* \sim 10^{0.5}$ to $\sim 10^{2} M_{\odot}$ pc $^{-2}$.  The relation then flattens at higher densities ($\Sigma_* \sim 10^2$ to $10^3 M_{\odot}$ pc$^{-2}$), reaching roughly solar metallicity. The {\it rms} residuals around the best fit relation are about $\pm$0.08 dex.
\item
The loci of the local values of $\Sigma_*$ and Z as measured in galaxies of different global stellar masses ($M_*$) systematically populate different parts of this relationship. As we move from low-mass to high-mass galaxies, the values for $\Sigma_*$ and $Z$ systematically increase: while all the data points lie along the same $\Sigma_*$ - Z relationship, low (high) mass disk galaxies are preferentially composed of regions of low (high) surface mass density and metallicity.
\item
Similarly, the loci of the local values of $\Sigma_*$ and Z as measured in galaxies of different global specific star-formation rates (sSFR) systematically populate different parts of this relationship. As we move from galaxies with high to low global sSFR, the values for $\Sigma_*$ and $Z$ systematically increase: while all the data points lie along the same $\Sigma_*$ - Z relationship, disk galaxies with high (low) global sSFR are preferentially composed of regions of low (high) surface mass density and metallicity.
\item
There are only weak systematic relationships between the residuals with respect to the best fit to the $\Sigma_*$ - Z relationship and either the global $M_*$ or sSFR. The largest systematic residuals are for the galaxies with very low $M_*$ ($\sim$ -0.08 dex below $\sim 10^{9.2} M_*$) or very high sSFR ($\sim$ -0.08 dex above sSFR $\sim 10^{-9.2}$ yr$^{-1}$).
\item
Using the best-fit to the $\Sigma_*$ - Z relation and the measured radial profiles of $\Sigma_{*}(r)$, we can accurately reproduce the observed radial metallicity gradients in disk galaxies spanning the range in global M$_*$ from $10^9$ to $10^{11}$ M$_{\odot}$. The worst disagreement ($\sim -0.1$ dex) was found in the inner regions ($< R_{50}$) in the lowest mass galaxies ($< 10^{9.5}$ M$_{\odot}$).
 \end{itemize}

The critical point is that a single local relationship between stellar surface mass density and gas-phase metallicity can simultaneously reproduce two different systematic properties of disk galaxies: their mass-metallicity relation and their radial metallicity gradients. These reproductions are not perfect over the full range of galaxy properties. In the next paper in this series we will explore the nature of possible 'secondary parameters' that could affect metallicity.

Put in the simplest possible terms, the results here show that the local gas-phase metallicity is very well determined by the local stellar surface-mass density: the outer disk of a massive galaxy and the inner regions of a low-mass galaxy (with the same density) have the same metallicity. This is a rather remarkable result in terms of the way in which galaxies assemble. 

\section*{Acknowledgements}

We thank R. Maiolino for useful discussion and revision on the manuscript. 
Funding for the Sloan Digital Sky Survey IV has been provided by
the Alfred P. Sloan Foundation, the U.S. Department of Energy Office of
Science, and the Participating Institutions. SDSS-IV acknowledges
support and resources from the Center for High-Performance Computing at
the University of Utah. The SDSS web site is www.sdss.org.

SDSS-IV is managed by the Astrophysical Research Consortium for the 
Participating Institutions of the SDSS Collaboration including the 
Brazilian Participation Group, the Carnegie Institution for Science, 
Carnegie Mellon University, the Chilean Participation Group, the French Participation Group, Harvard-Smithsonian Center for Astrophysics, 
Instituto de Astrof\'isica de Canarias, The Johns Hopkins University, 
Kavli Institute for the Physics and Mathematics of the Universe (IPMU) / 
University of Tokyo, Lawrence Berkeley National Laboratory, 
Leibniz Institut f\"ur Astrophysik Potsdam (AIP),  
Max-Planck-Institut f\"ur Astronomie (MPIA Heidelberg), 
Max-Planck-Institut f\"ur Astrophysik (MPA Garching), 
Max-Planck-Institut f\"ur Extraterrestrische Physik (MPE), 
National Astronomical Observatory of China, New Mexico State University, 
New York University, University of Notre Dame, 
Observat\'ario Nacional / MCTI, The Ohio State University, 
Pennsylvania State University, Shanghai Astronomical Observatory, 
United Kingdom Participation Group,
Universidad Nacional Aut\'onoma de M\'exico, University of Arizona, 
University of Colorado Boulder, University of Oxford, University of Portsmouth, 
University of Utah, University of Virginia, University of Washington, University of Wisconsin, 
Vanderbilt University, and Yale University.


\bibliographystyle{mnras}
\bibliography{mzsf_final}


\bsp	
\label{lastpage}
\end{document}